# Detecting Molecular Rotational Dynamics Complementing the Low-Frequency Terahertz Vibrations in a Zirconium-Based Metal-Organic Framework

Matthew R. Ryder,[1,2,3] Ben Van de Voorde,[4] Bartolomeo Civalleri,[5] Thomas D. Bennett,[6] Sanghamitra Mukhopadhyay,[2] Gianfelice Cinque,[3] Felix Fernandez-Alonso,[2,7] Dirk De Vos,[4] Svemir Rudić,[2] and Jin-Chong Tan[1,*]

[1]Multifunctional Materials & Composites (MMC) Laboratory, Department of Engineering Science, University of Oxford, Parks Road, Oxford OX1 3PJ, United Kingdom
[2]ISIS Facility, Rutherford Appleton Laboratory, Chilton, Didcot OX11 0QX, United Kingdom
[3]Diamond Light Source, Harwell Campus, Didcot, Oxford OX11 0DE, United Kingdom
[4]Centre for Surface Chemistry and Catalysis, KU Leuven, Arenbergpark 23, Leuven, B-3001, Belgium
[5]Department of Chemistry, NIS and INSTM Reference Centre, University of Turin, via Pietro Giuria 7, 10125 Torino, Italy
[6]Department of Materials Science and Metallurgy, University of Cambridge, Cambridge CB3 0FS, United Kingdom
[7]Department of Physics and Astronomy, University College London, Gower Street, London WC1E 6BT, United Kingdom

**Abstract:** We show clear experimental evidence of co-operative terahertz (THz) dynamics observed below 3 THz (~100 cm$^{-1}$), for a low-symmetry Zr-based metal-organic framework (MOF) structure, termed MIL-140A [ZrO(O$_2$C-C$_6$H$_4$-CO$_2$)]. Utilizing a combination of high-resolution inelastic neutron scattering and synchrotron radiation far-infrared spectroscopy, we measured low-energy vibrations originating from the hindered rotations of organic linkers, whose energy barriers and detailed dynamics have been elucidated via *ab initio* density functional theory (DFT) calculations. For completeness, we obtained Raman spectra and characterized the alterations to the complex pore architecture caused by the THz rotations. We discovered an array of soft modes with trampoline-like motions, which could potentially be the source of anomalous mechanical phenomena, such as negative linear compressibility and negative thermal expansion. Our results also demonstrate coordinated shear dynamics (~2.5 THz), a mechanism which we have shown to destabilize MOF crystals, in the exact crystallographic direction of the minimum shear modulus ($G_{min}$).

Terahertz (THz) vibrations play a central role in the identification of the structural flexibility of framework materials [1,2], and in the unravelling of their fundamental mechanistic dynamics [3,4]. These low-frequency vibrational modes were highlighted in relation to the structural destabilization of crystalline zeolites, and allowed for the Boson peak in the glassy state to be linked to framework features at the nanoscale, namely, connected ring moieties [3]. We have recently advanced the utility of THz vibrational modes to study the lattice dynamics of more complex 3-D hybrid framework materials, termed: metal-organic frameworks (MOFs) [1,5]. MOFs are crystalline inorganic-organic materials with long-range ordered porosity at the nanoscale, which have garnered immense scientific and technological interest for engineering innovative applications targeting carbon capture, smart sensors, opto- and micro-electronics, biomedicine, and energy conversion devices [6].

Because of the flexible nature of porous MOF architectures, we now know that collective vibrational motions are ubiquitous in the THz region. Rigorous *ab initio* simulations using density functional theory (DFT) can aid in explicitly linking the THz modes to rich physical phenomena, ranging from "gate-opening" and framework "breathing", to shearing deformations and other "soft modes"; for example in zeolitic MOFs [1]. We also recently reported how THz vibrations might be connected to the unusual elasticity in MOF mechanics, where co-operative cluster dynamics can offer insights into the origin of auxeticity (negative Poisson's ratios), an anomalous phenomenon predicted to occur in the copper-based paddle-wheel MOF, HKUST-1 [5].

The agreement, hitherto, between the limited sets of experiment and theory is of a reasonable quality for MOFs [1,2]. Measurement of THz vibrations in complex framework structures like MOFs is far from trivial, and an ongoing challenge is to accomplish high-resolution experimental data of all low-energy vibrational modes so that an improved comparison can be made against the idealized spectra, derived from DFT calculations.

Molecular rotors are a very topical area of nanoscience, as the engineering, building, and controlling of such molecular-scale "machines" is of high scientific and technological interest [7]. The rotational dynamics of several molecular crystals has been reported, such as difluorobenzene [8], and more recently of porphyrin-based double-decker complexes [9]. Rotary motions in porous molecular crystals and porous organic frameworks are currently of considerable interest, and the effects of guest inclusion on rotor dynamics have been explored [10]. The phenomenon of rotational motions is also present in 3-D MOF networks, where the rotors are organic linkers bridging the metal-coordination clusters (nodes) [11]. However, all of the previous literature has focused on the use of either nuclear magnetic resonance (NMR) or scanning tunnelling microscopy (STM) techniques; herein, we demonstrate the use of THz vibrations to detect the basic phenomenon of molecular rotors in a MOF structure and to study its accompanying lattice dynamics.

In this Letter, we report the low-frequency THz dynamics present in a porous Zr-based MOF, designated as MIL-140A [12], whose chemical structure is presented in Fig. 1. To gain a complete insight into the detailed cooperative framework dynamics, we employed inelastic neutron scattering (INS) (Fig. 2), synchrotron radiation far-infrared (SR-FIR) spectroscopy, and Raman spectroscopy (Fig. 3), all in conjunction with *ab initio* quantum mechanical calculations. To the best of our knowledge, the work demonstrates the first definitive MOF exemplar regarding a quantitative comparison of experimental and calculated spectra. This gives us the unique opportunity to unambiguously confirm three interesting physical phenomena predicted by quantum mechanical simulations, specifically at under 3 THz (<100 cm$^{-1}$): (i) hindered rotational dynamics, (ii) co-operative trampoline-like motions, and (iii) coordinated shearing dynamics.

To fully characterize and visualize the physical motions of each unique vibrational mode of MIL-140A, we computed the theoretical vibrational spectra at the $\Gamma$-point, using DFT via the CRYSTAL14 code [13]. The theoretical calculations were performed at the PBE exchange-correlation level of theory [14], adopting triple-zeta quality basis sets. We further applied a dispersion correction to the DFT functional (PBE-D) [15], to more accurately account for the van der Waals interactions. The PBE-D level of theory has been shown previously to give reliable agreement with experimental spectra [1]. However, we also calculated the spectra at the B3LYP/B3LYP-D level of theory to confirm the robustness of the calculations. Details regarding the methods of calculating the vibrational frequencies and their associated IR and Raman intensities [16] are included in the Supplemental Material [17].

We performed INS experiments at the ISIS Pulsed Neutron & Muon Source, to determine all of the vibrational modes, regardless of symmetry and hence avoiding any optical restrictions [18]. We have measured the broadband INS spectra on the TOSCA spectrometer [19], adopting the methodology we recently reported for studying zeolitic MOFs [1]. We subsequently conducted low-energy neutron scattering on the OSIRIS spectrometer [20]. This confirmed (a) the absence of any vibrational modes in the low wavenumber region under 16 cm$^{-1}$ (consistent with DFT theory), and (b) the optimized crystal structure at the PBE-D level of theory gives an improved agreement with the experimental neutron diffraction data (see Supplemental Material [17]).

The INS data were complemented by the SR-FIR measurements (Fig. 3), performed on the MIRIAM beamline [21] at the Diamond Light Source. We deployed an external liquid helium-cooled bolometer to significantly improve the signal-to-noise ratio and hence the detection of THz signals, permitting every IR-active peak to be observed experimentally. The reasoning for using both INS and SR-FIR methods was to allow us to have the advantage of a less complex spectra using SR-FIR spectroscopy, but at the cost of lower intensity signals in the critical THz region. The spectrum obtained via INS is much more complex, but with stronger signals in the THz region (as hydrogen has a large neutron cross section [18]).

Fig. 4 presents THz vibrations associated with the hindered rotor motion involving the BDC organic linkers (1,4-benzenedicarboxylate), specifically the $C_6H_4$ aromatic rings demonstrating out-of-plane torsional dynamics (see video clips in the Supplemental Material [17]). Interestingly, there is no specific mode that involves the simultaneous motion of all phenyl rings. Each rotational motion, therefore, can be classified as 'Type-A' or 'Type-B': depending if the linkers involved are the ones supporting the 4-node architecture of the framework (Type-A), or the linker in the middle reinforcing the 4-node unit (Type-B), marked in Fig. 1.

We discovered four modes encompassing rotor-like dynamics in MIL-140A (Figs. 2-4). These modes are mainly Raman-active, except the IR-active asymmetric Type-B rotor-like vibration located at 1.74 THz (58.1 cm$^{-1}$), which shows a simultaneous trampoline-like motion (*vide infra*). The other vibrations exhibiting the rotational motion are two modes: pinpointed at 1.34 and 1.48 THz (44.7 and 49.2 cm$^{-1}$), involving the asymmetric and symmetric Type-A rotor-like motions, respectively. Moreover, we identified a higher energy mode at 2.82 THz (94.2 cm$^{-1}$), exhibiting the symmetric Type-B rotations.

Of additional interest are the energy barriers to full phenyl rotation and the concurrent modification of the *solvent accessible volume* (SAV: quantifies accessible voids in the framework, calculated via PLATON [22]) at each degree of rotation. We calculated single-point energies at the PBE-D level of theory, for the full 180° twist ($\phi$) for both the symmetric Type-A and Type-B rotors. The energy barriers shown in Fig. 4 (inset) represent the simultaneous rotation of all Type-A or Type-B phenyl rings of one unit cell, containing 4 rings. The barrier for the Type-A rotation is approximately 1,130 meV, hence ~283 meV per aromatic ring. This is significantly lower than the ~537 meV value reported for MOF-5 [23] (because, for simplicity, ref. [23] rotated just one of the six phenyl rings present in the primitive cell).

We now characterize the modification of SAV on the full Type-A rotation, implicating a 2% increase from 26.4% at equilibrium ($\phi$ = 0° and 180°) to 28.4% at 20-30° and 90-100° (Fig. 4 inset). There is also a decrease to a minimum SAV of 24.6% at 160°. The varying level of SAV for the Type-A rotation can be directly linked to the changing morphology of the pore geometry (Fig. 5); although this is not as straightforward as "gate-open" or "gate-closed" configurations [1,24]. The multiple contributing factors come from the phenyl rings being positioned, so as to establish some accessible volume in-between the Type-A rings. Remarkably, the position of the rings resulting in the fully "gate-open" geometry confers the minimum SAV ($\phi_{min}$ =160°), as the pores are transformed into simply 1-D channels oriented in the *c*-axis [Fig. 5(c)].

Unlike the simultaneous Type-A rotations, the energy barrier and SAV for the Type-B rotation cannot be identified reliably, due to steric hindrance (*viz.* linkers overlapping). The full 180° rotation cannot be calculated, as the rings come into very close contact, as depicted in the negative amplitude of Fig. 4(d). The associated energies for the Type-B rotation are included in the Supplemental Material [17] for comparison.

The next type of motion of interest involves the organic linker moieties moving in a trampoline-like fashion (Fig. 6). Such deformation mechanisms have been suggested to yield negative thermal expansion (NTE) detected in the simpler cubic MOF structures: HKUST-1 [25] and MOF-5 [26]. Trampoline-like motions minimize the deformation of the carboxylate and aromatic groups, involving translation of the phenyl rings in and out of the plane of the equilibrium position of the linker. Shown in Figs. 2 & 3 are four such vibrational modes, discernible by the Type-A or -B linkers involved. As with the rotor-like motions, not all of the vibrations encompass every linker group. However, unlike the rotor-like motions, there are in fact two unique modes exhibiting the trampoline movement of every linker, simultaneously. They are both IR-active and located at 1.44 THz (48.2 cm$^{-1}$) and 1.61 THz (53.6 cm$^{-1}$); differing only in the direction of the Type-B trampoline motion which is simultaneous to the Type-A motion (i.e. the positive and negative amplitudes of the Type-B linkers for the 1.44 THz motion (see Supplementary Material [17]), are reversed in the mode located at 1.61 THz). The latter of the two modes, involving all of the linker moieties, is of significantly reduced IR intensity (26 times lower than the other IR-active trampoline motions). The other trampoline-like motions consist of: another IR-active mode at 1.44 THz (48.2 cm$^{-1}$), however only involving the Type-A linkers; and the final mode demonstrating trampoline motion is Raman-active located at 2.75 THz (91.8 cm$^{-1}$), which involves exclusively Type-B linkers. All of the trampoline-like dynamics can be better understood from the video clips supplied in the Supplemental Material [17].

Finally, we have observed a Raman-active mode at 2.47 THz (82.5 cm$^{-1}$), clearly visible in the INS spectra (Fig. 2), which could point to the mechanism responsible for structural destabilization via a coordinated shearing-type angular distortion in the ⟨010⟩ directions (video in Supplemental Material [17]). Notably, this particular vibrational signature has been predicted via DFT to have a lower intensity than the other collective motions discussed, also evidenced in the experimental INS data (Fig. 2). Of significant importance, we established that this shear motion arises in the exact direction that was previously reported to give the minimum value of shear modulus ($G_{min}$ = 3.2 GPa) for MIL-140A[29], substantiating our current observation. Crucially, this is the first report demonstrating the experimental observation of shear dynamics of a MOF structure. This is an important result because shear-type deformations have previously been speculated to be the mechanism behind framework destabilization of MOF systems [27], causing the loss of crystallinity via amorphization [28].

This Letter demonstrates a major step forward in the use of high-resolution THz spectroscopy, whereby experiments are validating the theory, and substantially increase our limited understanding of the complex physical mechanisms controlling the core functions of MOFs and related framework materials. New methodologies being achieved in the field mean that complete THz characterization of MOF vibrational dynamics is now possible. This work has opened the door to the accurate recognition and elucidation of remarkable physical phenomena, such as a multitude of rotor-like dynamics of linkers and associated trampoline-like vibrational motions, in addition to the first validation of the existence of low-energy shear dynamics that may trigger the destabilization of MOF crystals. Just as trampoline-like modes have been postulated to be the source of NTE in MOFs [25,26], our new findings will stimulate more intense efforts to search for and unravel new relationships enabled by previously unknown THz pathways behind unconventional mechanical properties. Such as negative linear compressibility (NLC) [30] and auxeticity [29,31], which are emergent topics in the vibrant field of framework materials.

### Acknowledgements

M. R. R. would like to thank the UK Engineering and Physical Sciences Research Council (EPSRC) for a DTA postgraduate scholarship and also an additional scholarship from the Science and Technology Facilities Council (STFC) CMSD Award 13-05. M. R. R. would also like to thank Rodolfo Fleury for helpful discussions and advice regarding Mathematica. We are also grateful to the ISIS Support Laboratories, especially Dr Marek Jura and Dr Gavin Stenning at the Materials Characterisation Laboratory (R53) for providing access to powder X-ray diffraction equipment. The experimental work was performed at large scale facilities through the ISIS Beamtime at TOSCA (RB1510531) and OSIRIS (RB1410426) and the Diamond Beamtime at B22 MIRIAM (SM10215). Via our membership of the UK's HEC Materials Chemistry Consortium, which is funded by EPSRC (EP/L000202), this work used the ARCHER UK National Supercomputing Service. We thank the Advanced Research Computing (ARC) facility (http://dx.doi.org/10.5281/zenodo.22558) at Oxford University

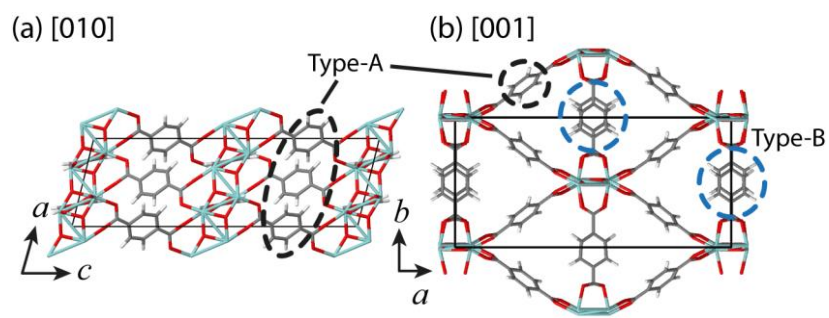

Figure 1. Three-dimensional (3-D) framework structure of MIL-140A [ZrO(BDC); BDC = $O_2C$-$C_6H_4$-$CO_2$] looking down the crystallographic (a) *b*-axis and (b) *c*-axis, respectively; it has the monoclinic *C*2/*c* space group. The inorganic building units are $ZrO_6$ coordination polyhedra, forming 1-D chains along the ⟨001⟩ zone axis. The black line represents one unit cell. Color scheme adopted: zirconium: light blue; carbon: gray; oxygen: red; hydrogen: white. Circled areas denote the Type-A and Type-B motions of the organic linkers.

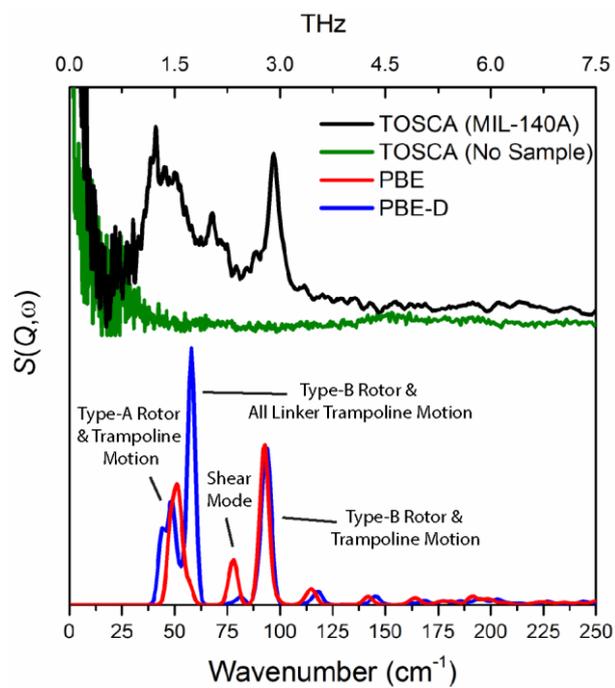

Figure 2. Comparison of the experimental (black) and theoretical INS spectra (red: PBE; blue: PBE-D) in the region of 0-250 cm$^{-1}$, measured at 5 K. Note that 1 THz ≈ 33.3 cm$^{-1}$. The experimental spectra excluding the MIL-140A sample is included to confirm the peaks are related to the framework vibrations.

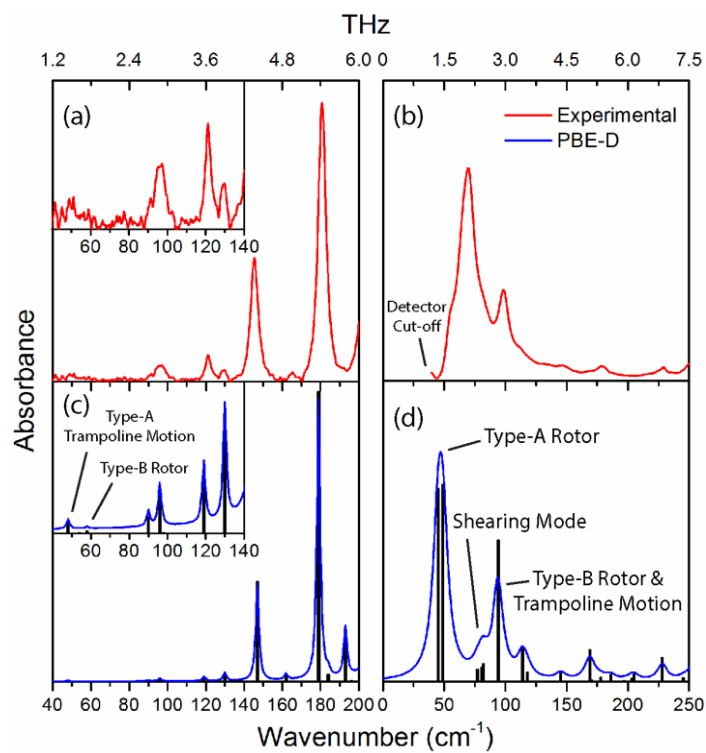

Figure 3. (a & c) SR-FIR and (b & d) Raman spectra in the region of 40-200 and 0-250 cm$^{-1}$ respectively. Comparison of experimental (red) and theoretical spectra (blue and black) for MIL-140A. The blue theoretical spectra have a full-width-half-maximum (FWHM) Lorentzian line fit applied to aid in comparison with the experimental data (2 cm$^{-1}$ for FIR and 5 cm$^{-1}$ for Raman). Theoretical Raman data were corrected for a laser frequency of 532 nm and a temperature of 293 K.

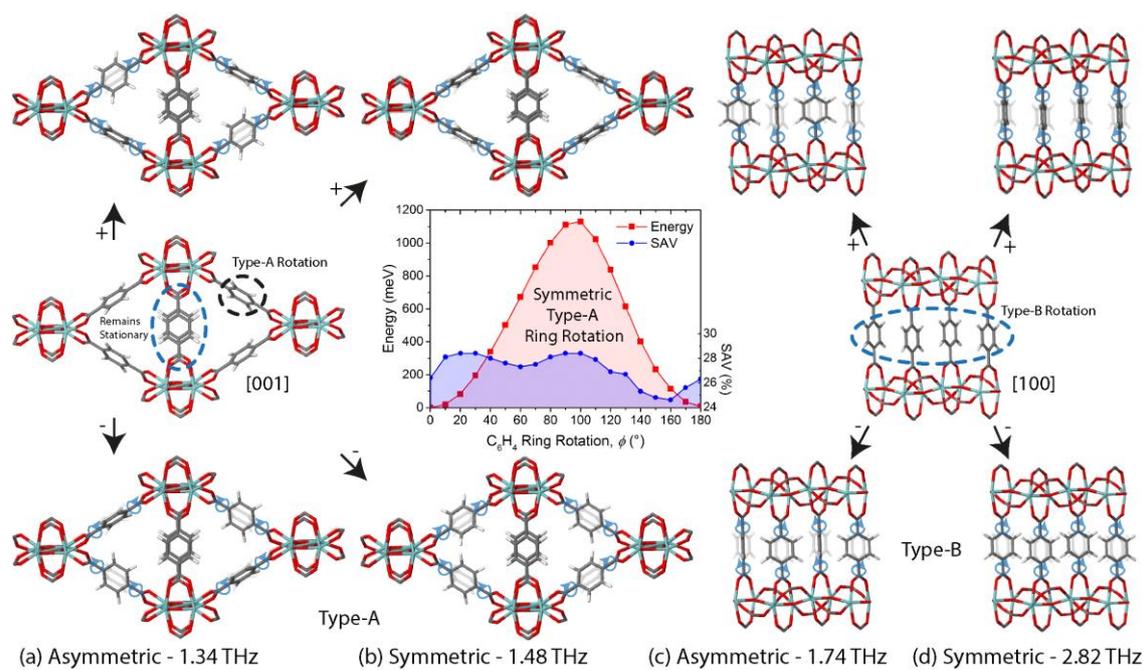

Figure 4. Rotational dynamics of MIL-140A determined from DFT calculations, in accordance with experimental observations (Figure 2). (a) Asymmetric and (b) symmetric Type-A rotor-like motions along with (inset) the energy barrier and change in SAV for the 180° ring rotations during the symmetric Type-A motion. (c) Asymmetric and (d) symmetric Type-B rotor-like motions. Blue circular arrowheads are used to mark the rotational directions.

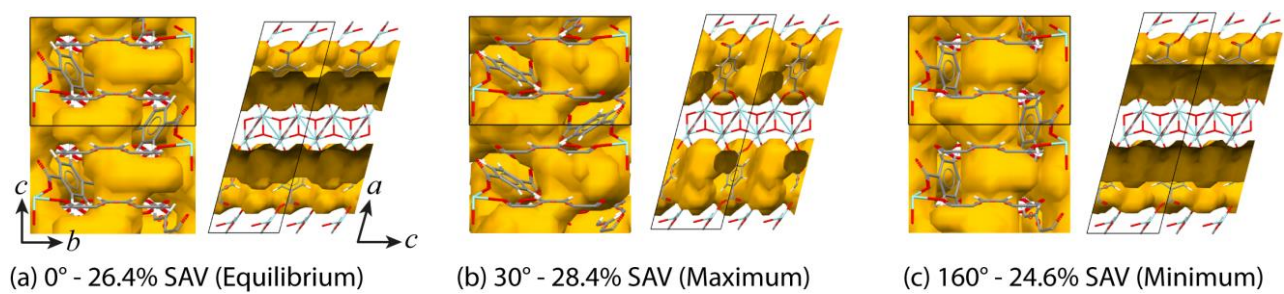

Figure 5. Modification of the solvent accessible volume (SAV) of the (a) equilibrium geometry, with the structures obtained when the Type-A aromatic rings are rotated symmetrically by (b) $\phi = 30°$, resulting in an increased SAV, and (c) $\phi = 160°$, resulting in a decreased SAV and formation of purely 1-D pore channels along the *c*-axis.

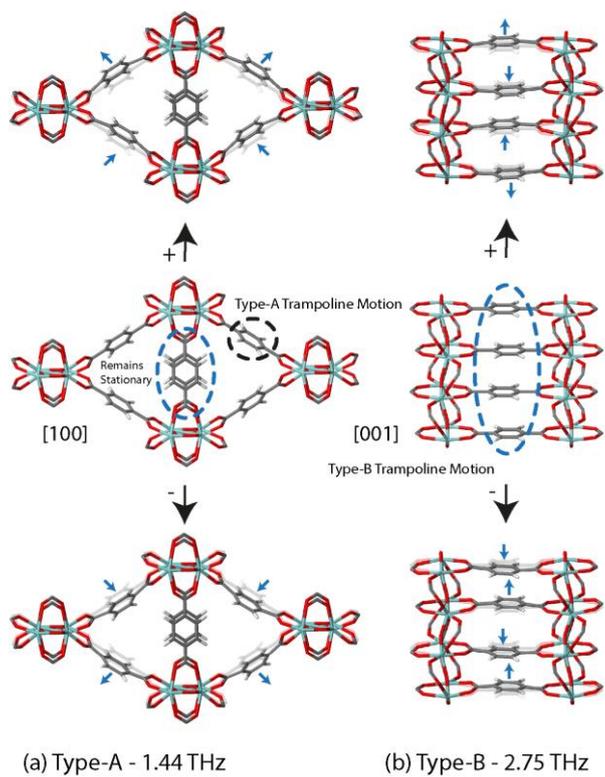

Figure 6. (a) Asymmetric Type-A and (b) asymmetric Type-B trampoline-like vibrational motions of MIL-140A.